\documentclass[prb,reprint,floatfix,dvipsnames]{revtex4-2}
\usepackage{fullpage, setspace, parskip, xcolor, breakcites, hyphenat}
\usepackage{graphicx, epstopdf, amsmath, amssymb, amsfonts, array,
	verbatim, url, subcaption, booktabs, multirow, mathptmx, etoolbox,siunitx-v2, etoolbox, tikz,stackengine}
\usepackage{lineno}
\usepackage[inline]{enumitem}
\usepackage[version=4]{mhchem}
\usepackage[utf8]{inputenc}
\usepackage[T1]{fontenc}
\usepackage{soul}
\usetikzlibrary{shadings}
\definecolor{srm}{HTML}{034da1}
\definecolor{cof}{RGB}{219,144,71}
\definecolor{pur}{RGB}{186,146,162}
\definecolor{greeo}{RGB}{91,173,69}
\definecolor{greet}{RGB}{52,111,72}
\definecolor{mplc0}{HTML}{1f77b4}
\definecolor{mplc1}{HTML}{ff7f0e}
\definecolor{mplc2}{HTML}{2e7d32}
\definecolor{mplc3}{HTML}{d32f2f}
\definecolor{mplc4}{HTML}{9467bd}
\definecolor{mplc5}{HTML}{8c564b}
\definecolor{mplc6}{HTML}{e377c2}
\definecolor{mplc7}{HTML}{7f7f7f}
\definecolor{mplc8}{HTML}{bcbd22}
\definecolor{mplc9}{HTML}{17becf}
\definecolor{Astral}{HTML}{1F77B4}
\definecolor{BG80}{HTML}{37474f}
\definecolor{Green}{HTML}{4CA540}
\definecolor{Red}{HTML}{F44336}
\usepackage[colorlinks=true, linkcolor=mplc0, urlcolor=mplc3, filecolor=mplc0, citecolor=mplc2,
	pdfstartview=FitV, pdftitle={}, pdfauthor={}, pdfsubject={}, pdfkeywords={}, pdfpagemode={},
	bookmarksopen=true, breaklinks]{hyperref}
\PassOptionsToPackage{hyphens}{url}

\newcommand{\figref}[1]{FIG. (\ref{#1})}

\newcommand{\tablref}[1]{table (\ref{#1})}
\newcommand{\dgh}{$|\Delta G_{\text{H}^\ast}|$}
\usepackage[export]{adjustbox}
\graphicspath{{figures}}
\date{\today}
\usepackage[textsize=tiny,backgroundcolor=mplc0!20,bordercolor=none]{todonotes}

\makeatletter
\def\@email#1#2{%
	\endgroup
	\patchcmd{\titleblock@produce}
	{\frontmatter@RRAPformat}
	{\frontmatter@RRAPformat{\produce@RRAP{*#1\href{mailto:#2}{#2}}}\frontmatter@RRAPformat}
	{}{}
}%
\makeatother
\begin{document}
\title{Tuning Catalytic Efficiency: Thermodynamic Optimization of Zr-Doped \ce{Ti3C2} and \ce{Ti3CN} MXenes for HER Catalysis}
\author{Shrestha Dutta}
\author{Rudra Banerjee}\email{rudrab@srmist.edu.in}
\affiliation{Department of Physics and Nanotechnology, SRM Institute of Science and Technology, Kattankulathur, Tamil
	Nadu, 603203, India}
\begin{abstract} 
	Hydrogen production via the Hydrogen Evolution Reaction (HER) is critical for sustainable energy solutions, yet the reliance on
	expensive platinum (Pt) catalysts limits scalability. Zirconium-doped (\ce{Zr}-doped) MXenes, such as \ce{Ti3C2} and
	\ce{Ti3CN}, emerge as transformative alternatives, combining abundance, tunable electronic properties, and high catalytic
	potential. Using first-principles density functional theory (DFT), we show that \ce{Zr} doping at 3\% and 7\% significantly
	enhances HER activity by reducing the work function to the optimal range of 3.5-4.5~eV and achieving near-zero Gibbs free
	energy (\dgh) values of 0.18-0.16~eV, conditions ideal for efficient hydrogen adsorption and desorption.
	Bader charge analysis reveals substantial charge redistribution, with enhanced electron accumulation at \ce{Zr} and \ce{N}
	sites, further driving catalytic performance. This synergy between optimized electronic structure and catalytic properties
	establishes \ce{Zr}-doped MXenes as cost-effective, high-performance alternatives to noble metals for HER. By combining
	exceptional catalytic efficiency with scalability, our work positions \ce{Zr}-doped MXenes as a breakthrough for green hydrogen
	production, offering a robust pathway toward renewable energy technologies and advancing the design of next-generation
	non-precious metal catalysts.
\end{abstract}
\maketitle
\section{Introduction}
The increasing global demand for energy, driven by industrial growth and population expansion, necessitates the urgent
development of sustainable and eco-friendly energy solutions \cite{Zhi_2017,Chu_2012}. Hydrogen, as a clean and renewable energy
carrier, has emerged as a promising solution to these challenges, aligning with global sustainability goals \cite{Glenk_2022,Fan_2021}.
Among the various hydrogen production methods, the Hydrogen Evolution Reaction (HER), which relies on water splitting,
has gained attention due to its cost-effectiveness and environmental benefits \cite{Shan_2021,Chen_2019}. The HER process
efficiently produces hydrogen gas with water as a renewable source and oxygen (\ce{O2}) as the sole byproduct, making
it an attractive route for green energy conversion \cite{Khan_2018}.

Electrocatalysts are pivotal to HER efficiency, accelerating the reaction kinetics and reducing energy barriers \cite{Shan_2021}.
Precious metal-based catalysts such as platinum (Pt) and ruthenium (Ru) are widely regarded for their superior catalytic
performance, achieving low overpotentials and high cathodic current densities \cite{Hansen_2021,Luo_2020}. However, the scarcity
and prohibitive costs of these materials pose significant obstacles to their large-scale deployment, emphasizing the need for
cost-effective alternatives \cite{Changqing_20,Zhang_2021}. As a result, extensive research efforts are directed towards
non-precious metal (NPM)-based electrocatalysts, which offer the potential to balance performance and cost \cite{Mistry_2016,Jayakumar_2017,Anjali_2019}.

Despite the progress made in NPM-based catalysts, challenges remain in optimizing their electrical conductivity and long-term
stability in aqueous environments, which are critical to maintaining high catalytic activity and durability \cite{Peng_2019,Bai_21,Shen_2021}.
Various classes of NPM-based catalysts, such as transition metal dichalcogenides (TMDs) \cite{Wu_2021}, metal
nitrides \cite{Jamil_2021}, and carbon-based materials like MXenes \cite{Zhang_2017}, have been explored for HER applications.
Among these, MXenes—two-dimensional (2D) transition metal carbides, nitrides, and carbonitrides—stand out for their promising
catalytic properties, including excellent electrical conductivity, large surface areas\cite{Gautam_2023,Li_2021}, and tunable
electronic structures\cite{Hart_2019}.

MXenes, represented by the general formula \ce{M_{n+1}X_nT_x}, where \ce{M} is a transition metal and \ce{X} is carbon or
nitrogen, possess surface terminations such as \ce{-F}, \ce{-O}, or \ce{-OH} \cite{Naguib_2014, Khazaei_2014,Ghidiu_2015}. These
materials have demonstrated high potential for HER due to their multiple active sites, robust mechanical properties, and
hydrophilicity, which help their interaction with water molecules \cite{IMANIYENGEJEH_2021,Wyatt_2021}. Moreover, the
tunability of their Fermi levels enhances their catalytic activity in HER by improving electron transfer efficiency
\cite{Gogotsi_2021,ASHRAF_2022}. Nevertheless, the catalytic performance of MXenes in HER is primarily governed by the Gibb's
free energy of hydrogen adsorption (\(\Delta G_{H^*}\)), with an optimal value close to zero being desirable for efficient
hydrogen adsorption and desorption during the reaction \cite{WANG_2022,Li_2021,Ma_2019,Bai_21}.

In addition to \dgh, the work function($\phi$), defined as ``the work needed to move an electron from a surface to a
point in vacuum sufficiently far away from this surface \cite{wf_vasp} has also emerged as a good catalytic descriptor in recent time
\cite{Radinger_2022, Calle_Vallejo_2013}. The value of $\phi$ too less($\leqslant 3eV$) or too high ($\geqslant 6eV$)hinders either the Volmer or the
Heyrovsky/Tafle kinematics. The "habitable zone" of the $\phi$ is found out to be in the range of 3.5-4.5eV\cite{Radinger_2022,
	Chen_2024}.




Titanium carbide-based MXenes \ce{Ti3C2} have garnered significant attention as potential catalysts for the
hydrogen evolution reaction (HER) due to their tunable electronic structure, high surface area, and excellent hydrophilicity
\cite{Pang_2019, SREEDHAR_2023}. However, their catalytic performance is limited by several shortcomings, including suboptimal
hydrogen adsorption energy \dgh, insufficient active site density, and sluggish electron transfer kinetics
\cite{AZADI2023106136}. The metallic nature of \ce{Ti3C2} often leads to hydrogen adsorption that is either too strong or
too weak, which hinders the HER process. Additionally, the surface terminations typically present on \ce{Ti3C2}, such as
\ce{-OH} \ce{-F}, and \ce{-O}, introduce localized electronic states that are unfavorable for catalytic activity, thereby increasing the
overpotential and reducing turnover frequency \cite{Ran_2017, Jiang_2022}.

Doping has emerged as a promising strategy to address these limitations by modulating the electronic properties and surface
chemistry of \ce{Ti3C2}. Transition metal doping (e.g., Fe, Co, or Ni) can adjust the $d$-band center of \ce{Ti3C2}, optimizing
\dgh~ and enhancing its catalytic activity\cite{WANG2021114565}. Non-metal doping, such as nitrogen or
boron, can further enhance charge redistribution and stabilize defect states, thereby improving electron transfer\cite{WANG_2022}. For instance, Ni-doped Ti$_3$C$_2$ has demonstrated significantly reduced overpotential and enhanced HER
efficiency due to improved hydrogen adsorption and charge transport properties\cite{GOTHANDAPANI_2024, WANG2021114565}.

In this context, our study explores \ce{Zr}-doped \ce{Ti}-based MXenes as potential HER catalysts, aiming to overcome the
limitations of pristine \ce{Ti3C2} and \ce{Ti3CN} monolayers. By introducing Zr dopants at concentrations of 3\% and 7\%, we
investigate the impact of doping on the electronic structure, catalytic activity, and \dgh~ values of these
MXenes. Additionally, we examine the effects of hydrogen coverage on unterminated MXene monolayers to further elucidate the
catalytic mechanisms. Our results demonstrate that Zr-doping can significantly enhance the HER catalytic performance of Ti-based
MXenes, offering a pathway towards the development of cost-effective and efficient electrocatalysts for sustainable hydrogen
production.

\section{Computational Details}
All calculations were performed within the framework of density functional theory (DFT)\cite{Kurth_2005} using the
projector-augmented wave (PAW) method \cite{Blochl_1994}, as implemented in the Vienna \textit{Ab-initio} Simulation Package
(VASP) \cite{Kresse_1994,Kresse_1996,KresseG_96,Kresse_1999}. The exchange-correlation effects were treated using the
Perdew-Burke-Ernzerhof (PBE) functional under the generalized gradient approximation (GGA) \cite{Ernzerhof_1999}. To ensure the
accurate representation of valence electron-ion interactions, the plane-wave pseudopotential approach was employed, with a
plane-wave cutoff energy set to 520 eV.

The primary focus of this work was on monolayer Ti-based MXenes, both in pristine and Zr-doped configurations. A vacuum layer of
approximately 16 \AA\ was introduced along the $z$-axis to eliminate spurious interactions between periodic images of the layers.
For structural relaxations, a convergence criterion of $1.0 \times 10^{-3}$ eV/\AA\ was adopted, ensuring precise ionic
positions.

A $\Gamma$-centered Monkhorst-Pack \cite{Monkhorst_1976} $k$-point mesh of $14 \times 14 \times 2$ was used for the MXene unit
cell, with an equivalent $k$-point density applied to $3 \times 3 \times 1$ supercells. The choice of this $k$-point sampling
ensured adequate Brillouin zone integration for both pristine and doped systems. To capture the strong Coulombic interactions in
the $d$-orbitals, the Hubbard $U$ correction was applied using $U_{\text{eff}} = 3.0$ eV for Ti \cite{Sakhraoui_2022} and
$U_{\text{eff}} = 2.85$ eV for Zr \cite{Gebauer_2023}. The total energy minimization and convergence of force components were
carefully monitored to ensure high accuracy in all simulations.

The thermodynamic stability of the MXene structures was evaluated by calculating the formation energy ($FE$) using the standard
expression:
\begin{equation}
	FE(\ce{M3C2}) = E(\ce{M3C2}) - 3\mu_{\ce{M}} - 2\mu_{\ce{C}},
\end{equation}
where $E(\ce{M3C2})$ represents the total energy of the MXene compound, and $\mu_{\ce{M}}$ and $\mu_{\ce{C}}$ are the chemical
potentials of the metal and carbon atoms respectively, in their standard reference states.

The Volmer-Heyrovsky or the Volmer-Tafle reaction path of the HER ($2\ce{H}^+2e^-\rightarrow \ce{H2}$)
\begin{equation}
	\begin{aligned}
		H^++e^-        & \rightarrow H^\ast         & \qquad \text{Volmer Step}    \\
		H^\ast+H^++e^- & \rightarrow \ce{H_2}+\ast  & \qquad \text{Heyrovsky Step} \\
		2H^\ast        & \rightarrow \ce{H_2}+2\ast & \qquad \text{Tafel Step}
	\end{aligned}
\end{equation}
is largely controlled by the free energy of \ce{H} absorption, or the Gibbs free energy ($\Delta G_{H^*}$)\cite{WANG_2022,Li_2021}.
The necessary but insufficient condition for a material as ``good" HER catalyst is $|\Delta G_{H^\ast}|\approx 0$.
The free energy
change for the intermediate H$^*$ adsorption state is given by:
\begin{equation}
	\begin{aligned}
		\Delta G_{\ce{H^*}} & = \Delta E_{\ce{H^*}} + \Delta ZPE_{\ce{H^*}} - T \Delta S_{\ce{H^*}}      \\
		\Delta E_{\ce{H^*}} & = E_{\text{Layer} + \ce{H^*}} - E_{\text{Layer}} - \frac{1}{2}E_{\ce{H2}},
	\end{aligned}
\end{equation}
where \(\Delta E_{\ce{H^*}}\) is the adsorption energy of hydrogen, $E_{\text{Layer} + \ce{H^*}}$ and $E_{\text{Layer}}$ represent
the total energies of the hydrogen-adsorbed and clean MXene surfaces, respectively, and $E_{\ce{H2}}$ is the total energy of
molecular hydrogen in the gas phase.

\begin{figure}[htpb!]
	\centering
	\begin{subfigure}[b]{0.32\columnwidth}
		\centering
		\includegraphics[width=\columnwidth]{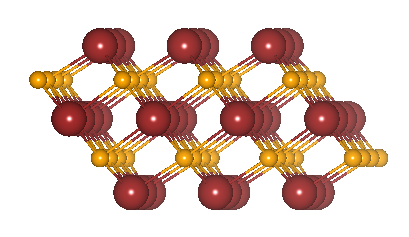}
		\caption{}
		\label{fig:tc_mono}
	\end{subfigure}
	\begin{subfigure}[b]{0.32\columnwidth}
		\centering
		\includegraphics[width=\columnwidth]{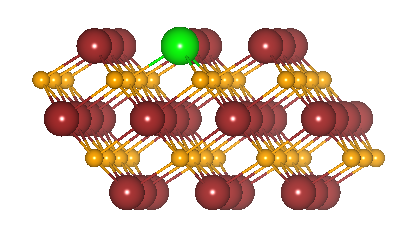}
		\caption{}
		\label{fig:tzc_1}
	\end{subfigure}
	\centering
	\begin{subfigure}[b]{0.32\columnwidth}
		\centering
		\includegraphics[width=\columnwidth]{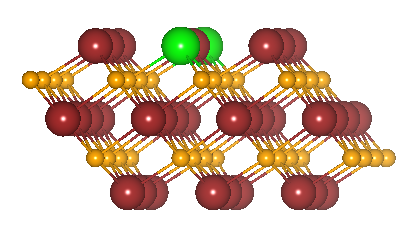}
		\caption{}
		\label{fig:tzc_2}
	\end{subfigure}
	\centering
	\begin{subfigure}[b]{0.32\columnwidth}
		\centering
		\includegraphics[width=\columnwidth]{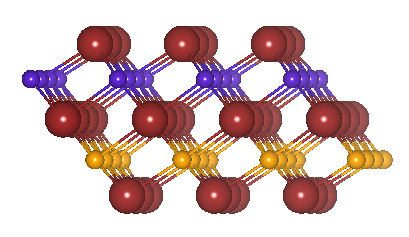}
		\caption{}
		\label{fig:tcn_mono}
	\end{subfigure}
	\begin{subfigure}[b]{0.32\columnwidth}
		\centering
		\includegraphics[width=\columnwidth]{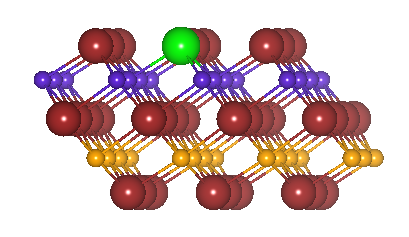}
		\caption{}
		\label{fig:tzcn_1}
	\end{subfigure}
	\begin{subfigure}[b]{0.32\columnwidth}
		\centering
		\includegraphics[width=\columnwidth]{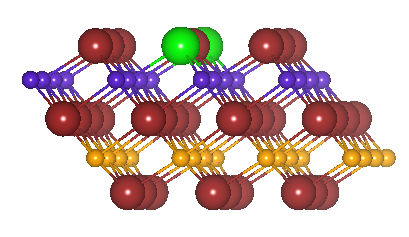}
		\caption{}
		\label{fig:tzcn_2}
	\end{subfigure}
	\caption{Monolayer structures of ({\subref{fig:tc_mono}}) \ce{Ti3C2}; ({\subref{fig:tzc_1}})
	\ce{(Ti_{.97}Zr_{.03})3C2}; ({\subref{fig:tzc_2}}) \ce{(Ti_{.93}Zr_{.07})3C2}; ({\subref{fig:tcn_mono}}) \ce{Ti3CN};
	({\subref{fig:tzcn_1}}) \ce{(Ti_{.97}Zr_{.03})3CN}; and ({\subref{fig:tzcn_2}}) \ce{(Ti_{.93}Zr_{.07})3CN}. Ti, Zr,
	C and N atoms are depicted as maroon, green, orange and violet balls, respectively.}
	\label{fig:monolayer}
\end{figure}
The zero-point energy correction (\(\Delta ZPE_{\ce{H^*}}\)) accounts for the quantum vibrational effects of the adsorbed hydrogen
atom, with values ranging from 0.01 to 0.04 eV for metal surfaces\cite{Sahoo_2022}. 
($T \Delta S_{\ce{H^*}}$), approximately 0.4 eV at room temperature, reflects the entropy difference between the adsorbed
hydrogen and the gas-phase \ce{H2} molecule \cite{Sen_2020,Huang_20}. Based on Nørskov’s volcano plot, optimal HER catalysts
exhibit $|\Delta G_{\ce{H^*}}| \leq 0.3$ eV \cite{Sen_2020}, a criterion we employed to identify promising electrocatalysts.

The work function has been calculated using
\begin{equation*}
	\phi=e\phi_{vaccum}-E_F
\end{equation*}
where $\phi_{vaccum}$ is the potential far away from the surface.
\section{Results and discussion}
\figref{fig:monolayer} shows the schematic of the dopped and undopped structures used in this study. \figref{fig:tc_mono} and
\figref{fig:tcn_mono} shows the \ce{T3C2} and \ce{T3CN} structures respectively. \figref{fig:tzc_1} and \figref{fig:tzc_2} are
showing 0.3\% and 0.7\% doping of \ce{Zr} at \ce{Ti} site of \ce{T3C2}(Those structures will be referred as \ce{TZ3C} and
\ce{TCZ7} for brevity in the rest of this manuscript). \figref{fig:tzcn_1} and \figref{fig:tzcn_2} are showing 0.3\% and 0.7\%
doping of \ce{Zr} at \ce{Ti} site of \ce{T3CN}, and will be referred as \ce{TZ3CN} and \ce{TZ7CN} in the rest of the manuscript.
All the dopped systems are found to be thermodynamically stable (2nd column of \tablref{tab:table0}). Note
that increased \ce{Zr} doping in \ce{T3C2} makes the system unstable, but due to higher electronegativity of \ce{N} compared to
\ce{C}, \ce{Zr} doped \ce{Ti3CN} shows the increase in stability. The result aligns with the reported energy above the hull
values of 0.05 eV/atom for \ce{Ti3C2}\cite{mp-ti3c2} and 0.02 eV/atom for \ce{Zr3N2}\cite{mp-zr3n2}, indicating enhanced
stability of the \ce{Zr}-dopped system.

\begin{figure*}[htpb]
	\centering
	\begin{subfigure}[b]{0.32\textwidth}
		\centering
		\includegraphics[width=\textwidth]{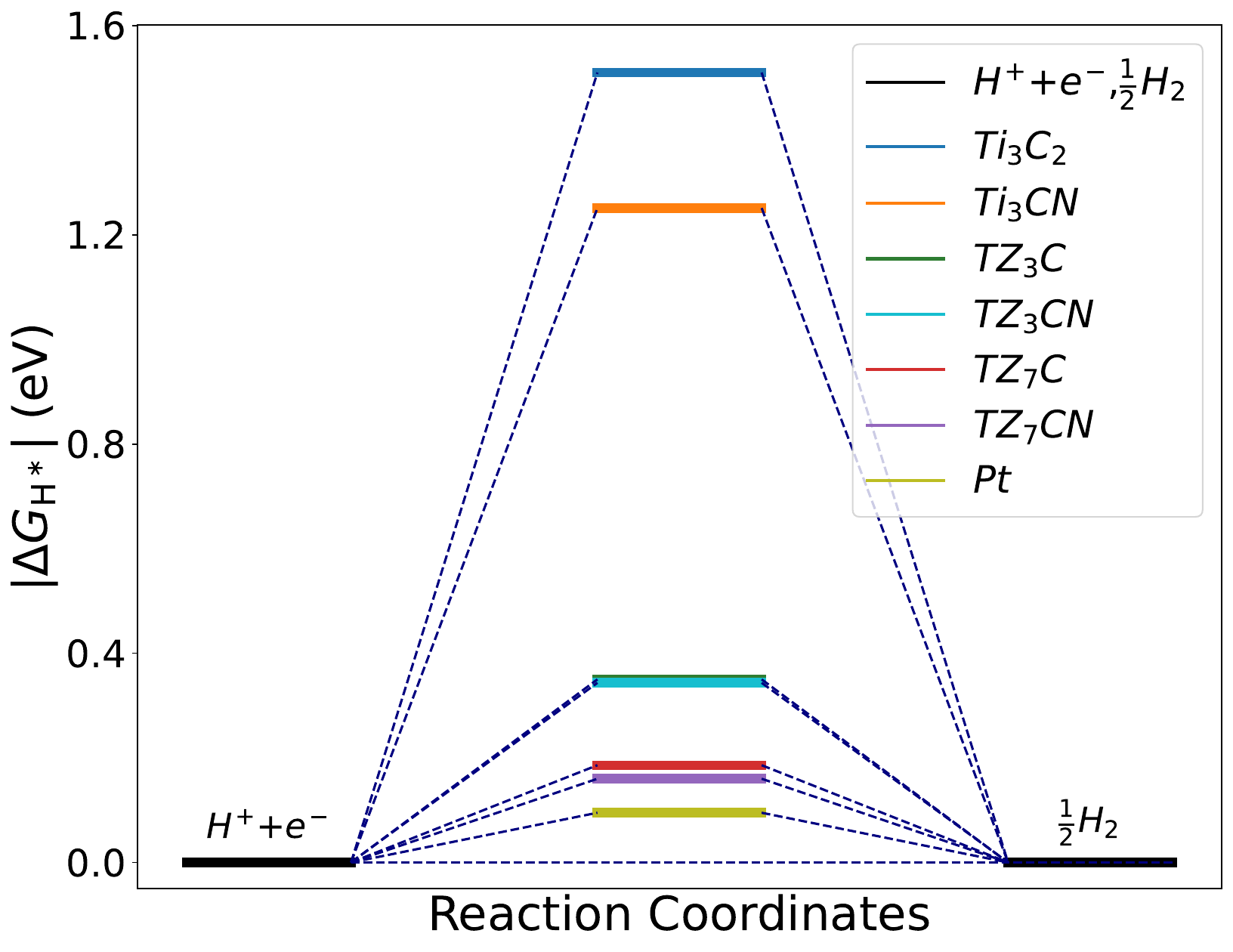}
		\caption{}
		\label{fig:gibbs}
	\end{subfigure}
	\begin{subfigure}[b]{0.32\textwidth}
		\centering
		\includegraphics[width=\textwidth]{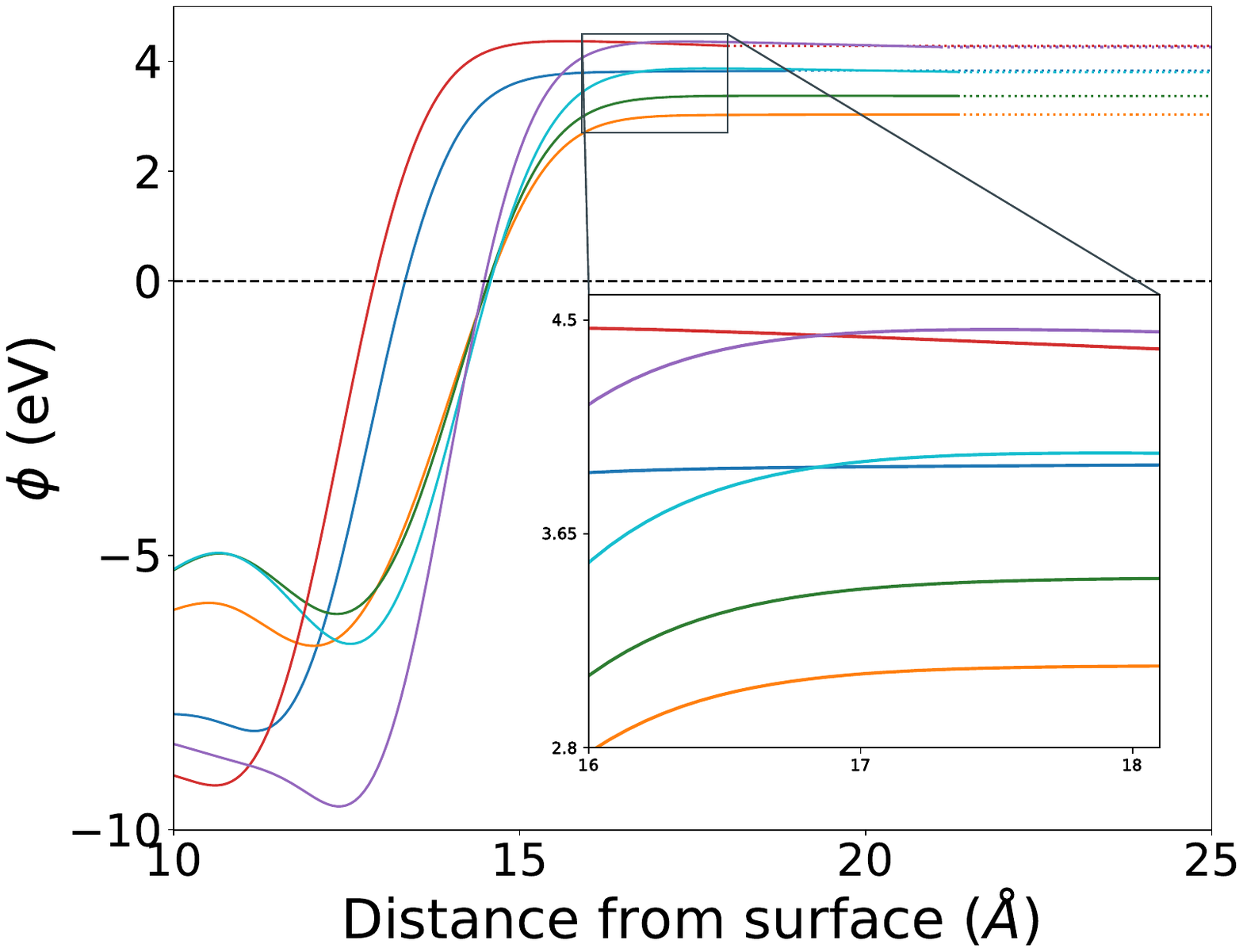}
		\caption{}
		\label{fig:wf}
	\end{subfigure}
	\begin{subfigure}[b]{0.32\textwidth}
		\includegraphics[width=\textwidth]{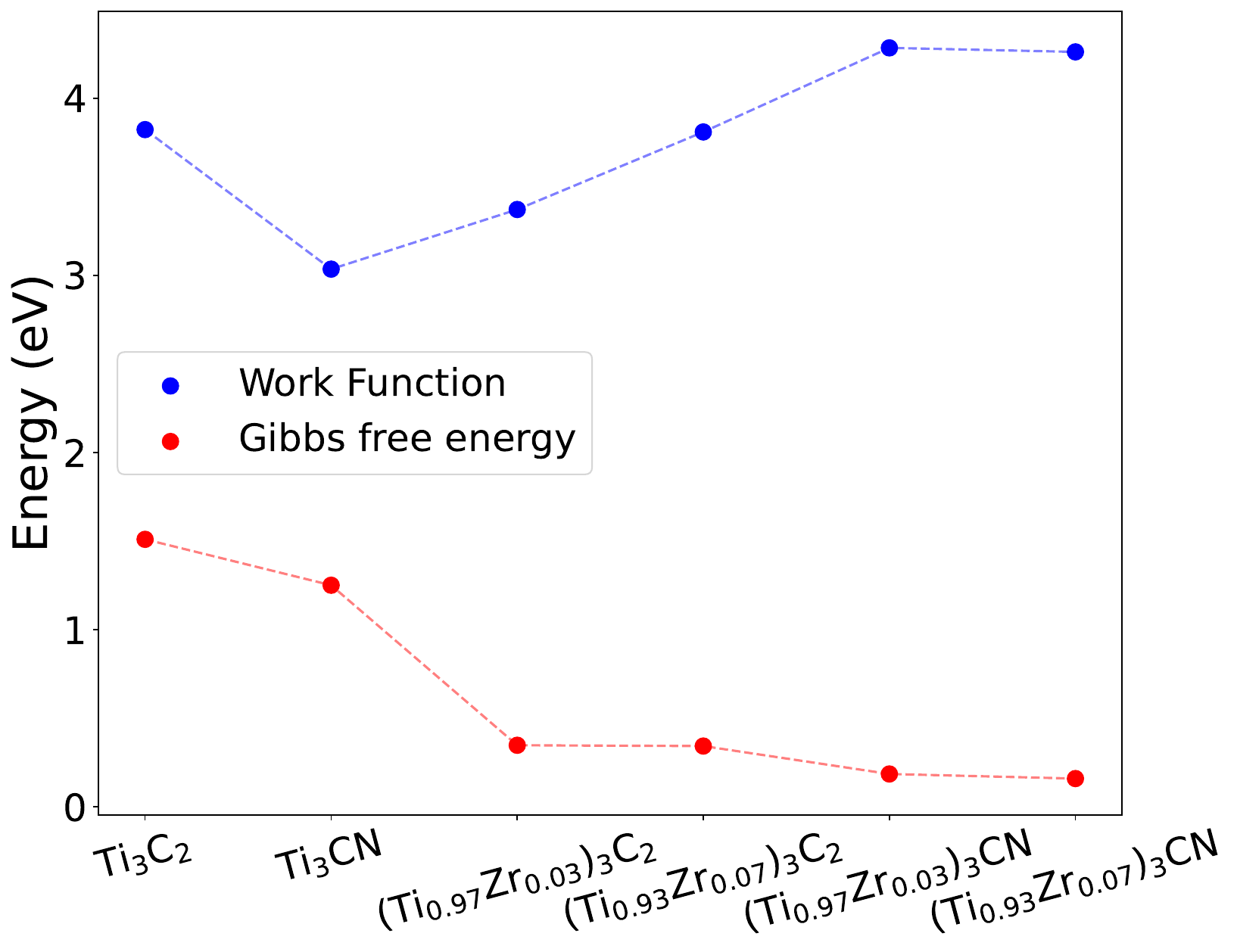}
		\caption{}
		\label{fig:corrwfgibbs}
	\end{subfigure}
	\caption{(\subref{fig:gibbs})\dgh~ for the MXene systems studied, with reference to Pt;({\subref{fig:wf}}) calculated work
		function normal to the surface. The dotted part of each curves are the constant line for comparison, and not actually
		calculated. Inset shows the zoomed in figure for better visualization. The line color indicates same systems as in
		(\subref{fig:gibbs}); and ({\subref{fig:corrwfgibbs}}) correlation between $\phi$ and \dgh.}
	\label{fig:wfandgibbs}
\end{figure*}

The catalytic performance of each of the above systems are evaluated through calculations of the \dgh~ as discussed
above. \figref{fig:gibbs} shows the \dgh~ of the systems studied. The results shows systematic decrease of \dgh~ with higher
concentrations of \ce{Zr}.
For pristine \ce{Ti3C2}, the \(\Delta G_{\ce{H^*}}\) value is 1.510 eV, while \ce{Ti3CN} shows a slightly lower value of
1.251 eV, both of which indicate suboptimal HER activity. In contrast, the introduction of Zr into the MXene lattice
significantly reduces \dgh, with \ce{TZ7C} and \ce{TZ7CN} exhibiting values
of 0.185 eV and 0.160 eV, respectively. These values are remarkably close to the ideal range for HER, suggesting that Zr-doping
dramatically enhances the catalytic efficiency of the MXenes. The values are tabulated in \tablref{tab:table0} for comparison.


To elucidate the origin of the significant reduction in \dgh~ observed with \ce{Zr} doping, we performed comprehensive electronic
structure calculations to determine the work function ($\phi$) and Bader charge distribution.
The work function is a critical parameter in determining the efficiency of materials for the HER. \figref{fig:wf} shows the
$\phi$ with respect to the surface. We observe significant variations in $\phi$ with doping concentration. The pristine MXenes,
\ce{Ti3C2} and \ce{Ti3CN}, exhibit relatively higher work functions, limiting their HER performance. Upon doping with 3\% Zr,
i.e. \ce{TZ3C} and \ce{TZ3CN}, show reduced work functions, nearing the optimal range, suggesting enhanced catalytic activity.
Increasing the Zr content to 7\% in \ce{TZC7} and \ce{TZ7CN} further takes the work function squarely within the desired range.
This systematic change demonstrates that \ce{Zr} doping effectively tunes the electronic properties of MXenes, enhancing their
HER efficiency.  Furthermore, the correlation between \dgh~ and the work function, illustrated in \figref{fig:corrwfgibbs}, reveals a robust linear relationship, underscoring the direct impact of electronic structure
modifications on catalytic performance. These findings align with prior studies that have demonstrated a strong link between the
electronic properties of MXenes and their catalytic activity\cite{Jiang_2022}. The observed trend suggests that Zr doping not
only tunes the electronic environment but also optimizes the catalytic properties of the material, making Zr-doped Ti-based
MXenes a promising candidates for HER applications.

\begin{figure*}[htpb]
	\begin{tikzpicture}
		\node[rotate=90] at (0,100){\small{DOS (states/eV (arb. unit))}};
	\end{tikzpicture}
	\centering
	\begin{subfigure}[t]{0.30\textwidth}
		\centering
		\phantomsubcaption
		\includegraphics[width=\textwidth]{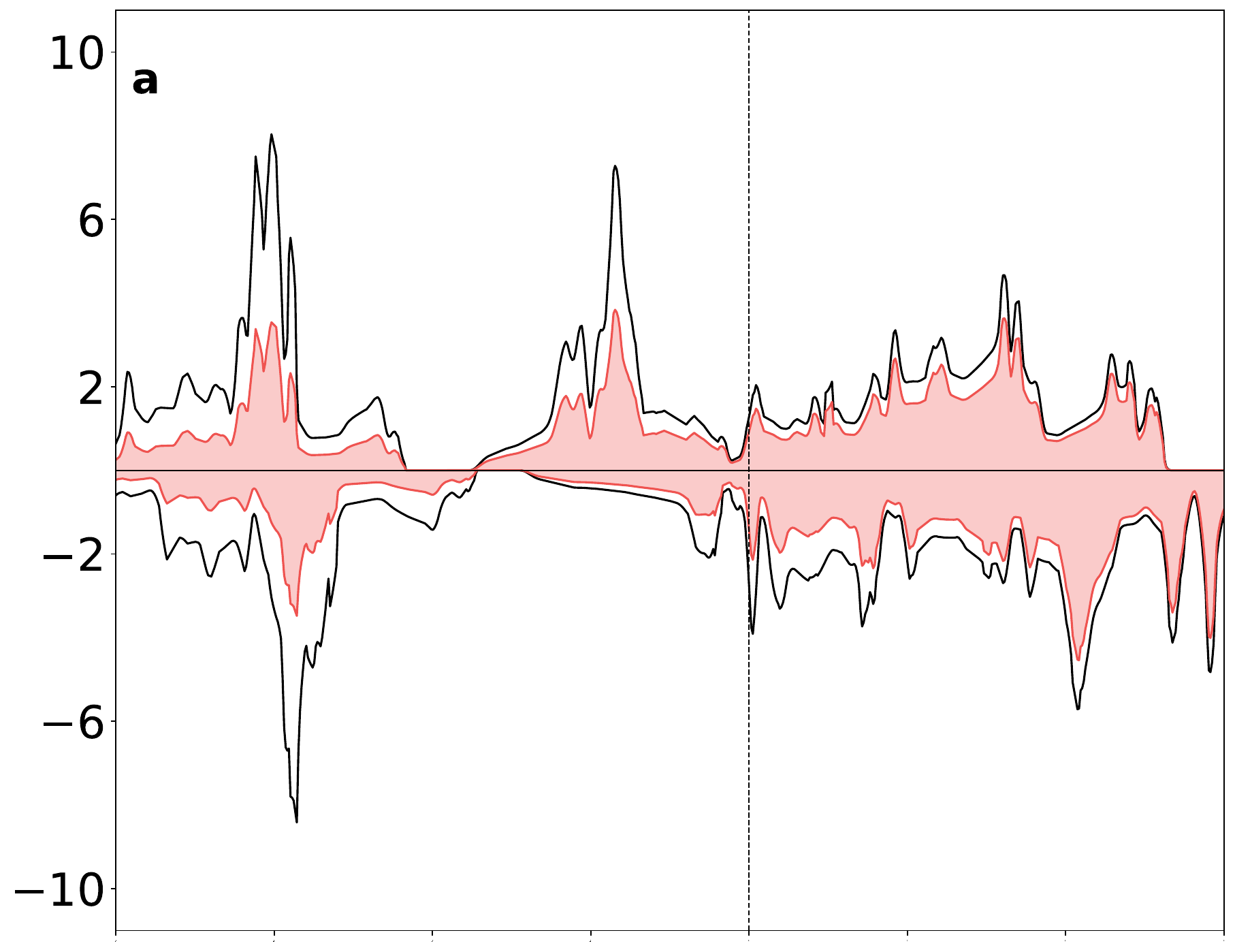}
		\label{fig:tc_cal_u}
	\end{subfigure}
	\hspace{-.5em}
	\begin{subfigure}[t]{0.30\textwidth}
		\centering
		\phantomsubcaption
		\includegraphics[width=\textwidth]{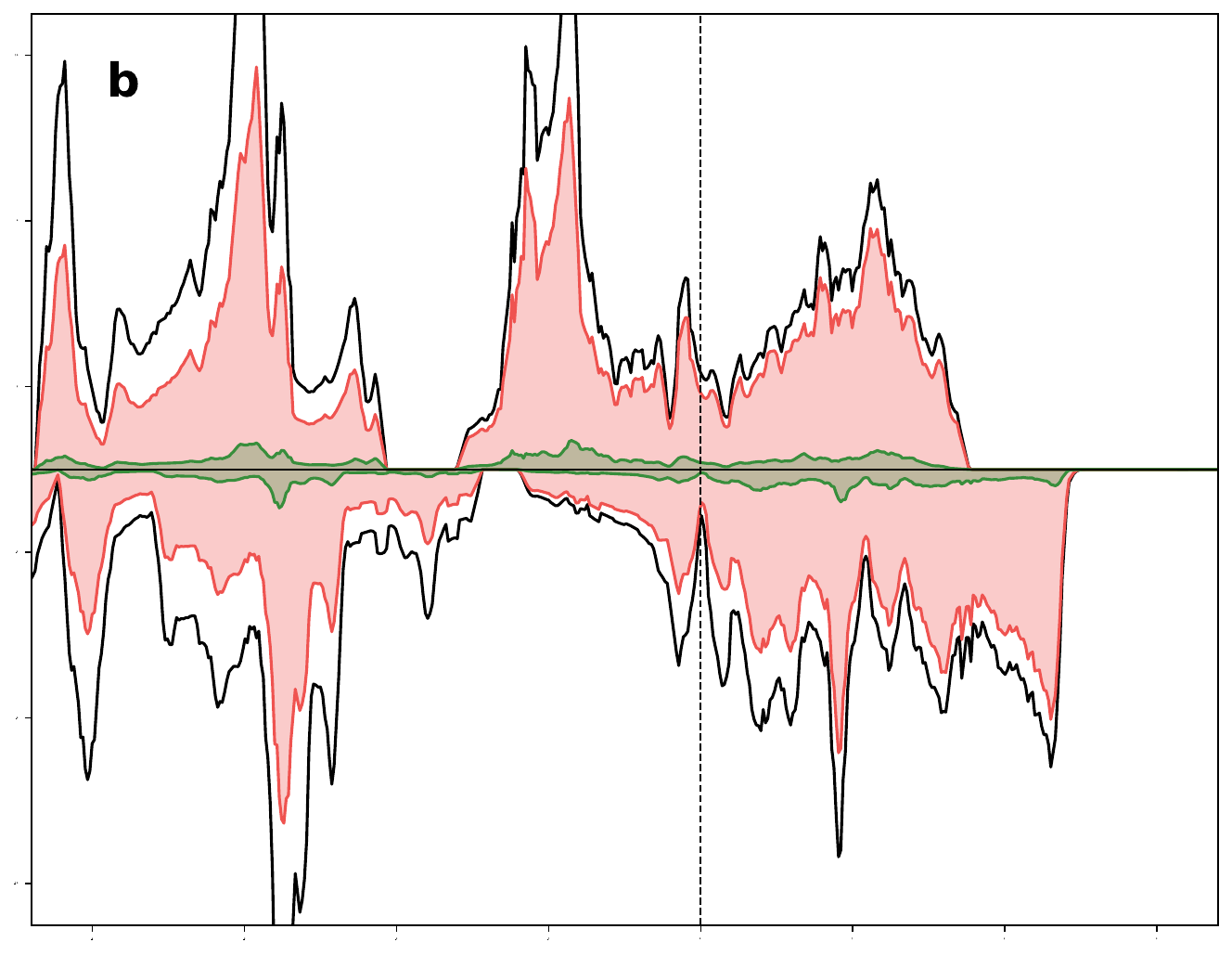}
		\label{fig:tzc_cal_u}
	\end{subfigure}
	\hspace{-.5em}
	\begin{subfigure}[t]{0.30\textwidth}
		\centering
		\phantomsubcaption
		\includegraphics[width=\textwidth]{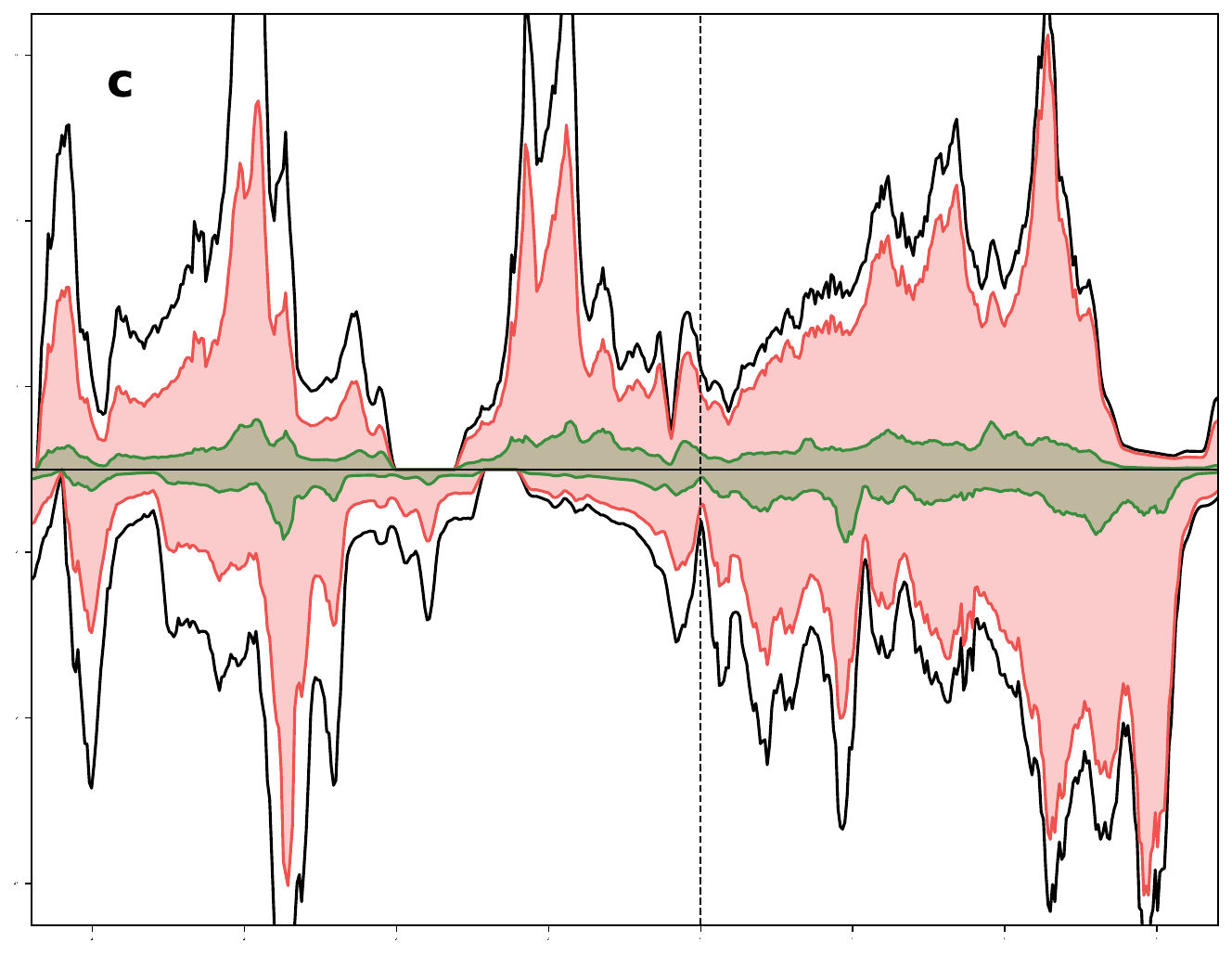}
		\label{fig:tzc2_cal_u}
	\end{subfigure}
	\par\vspace{-1.5em}
	\begin{tikzpicture}
		\node[rotate=90] at (0,100){\small{DOS (states/eV (arb. unit))}};
	\end{tikzpicture}
	\begin{subfigure}[t]{.30\textwidth}
		\centering
		\includegraphics[width=\textwidth]{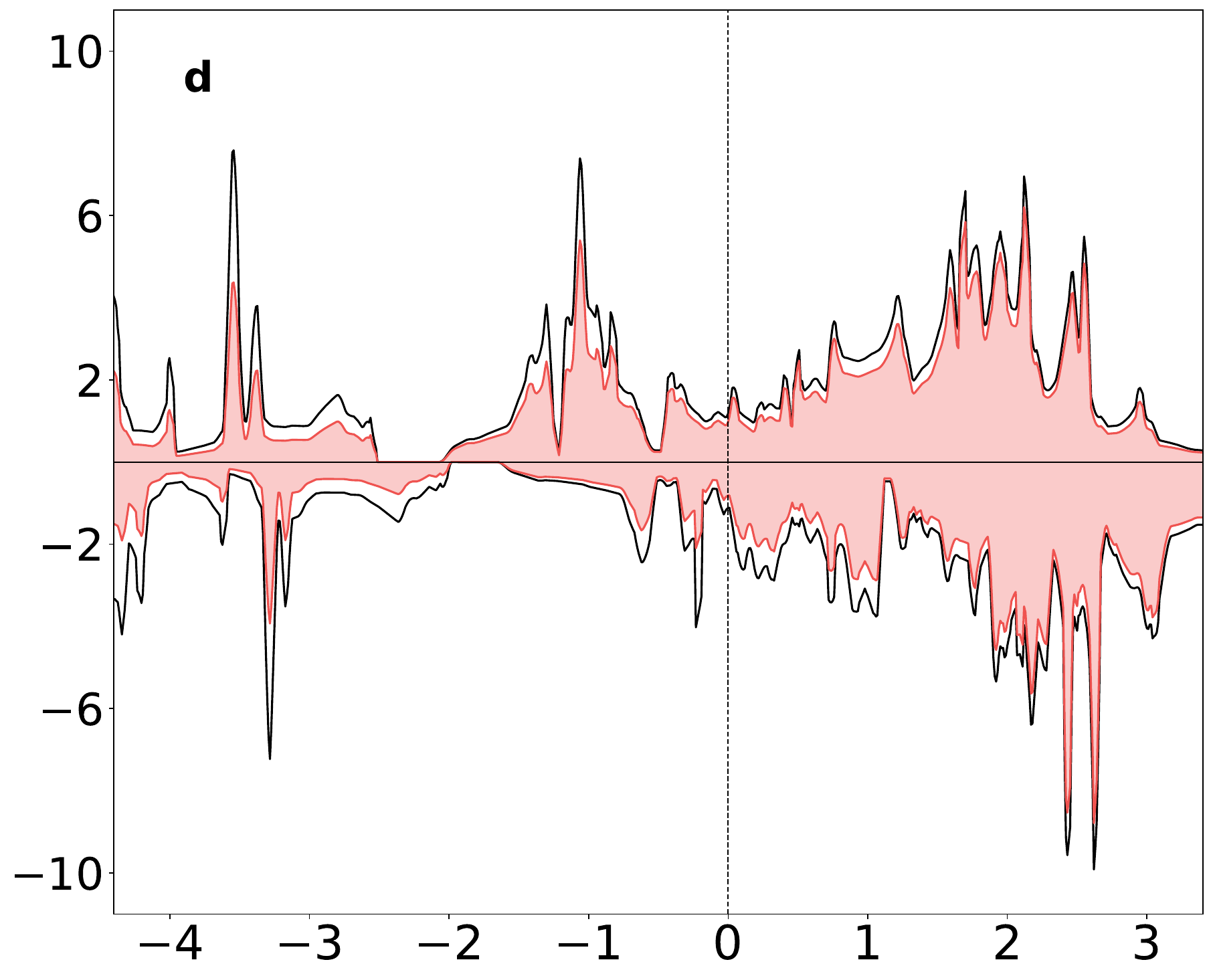}
		\phantomsubcaption
		\label{fig:tcn_cal_u}
	\end{subfigure}
	\hspace{-.5em}
	\begin{subfigure}[t]{0.30\textwidth}
		\centering
		\includegraphics[width=\textwidth]{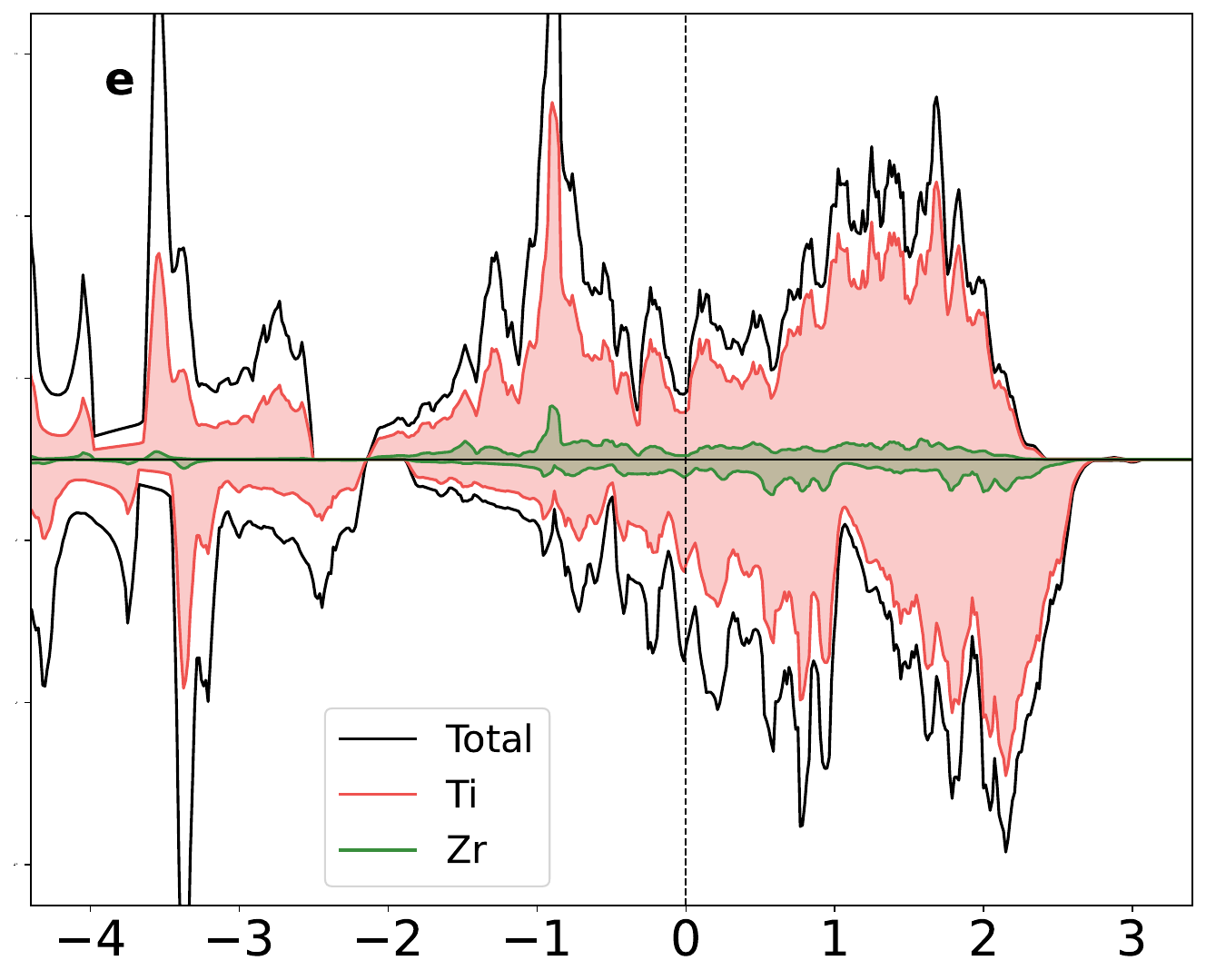}
		\phantomsubcaption
		\label{fig:tzcn_cal_u}
	\end{subfigure}
	\hspace{-.5em}
	\begin{subfigure}[t]{0.30\textwidth}
		\centering
		\includegraphics[width=\columnwidth]{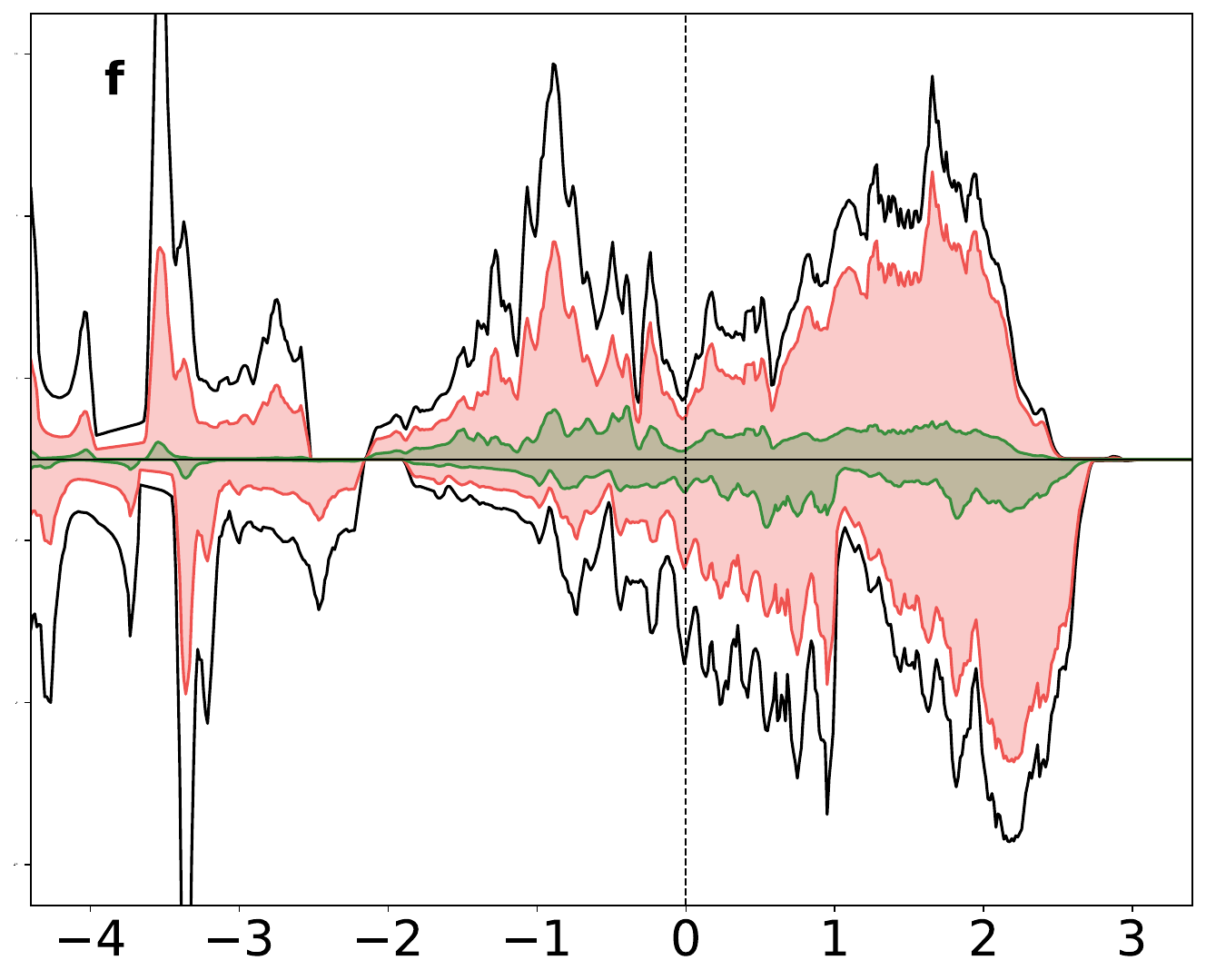}
		\phantomsubcaption
		\label{fig:tzcn2_cal_u}
	\end{subfigure}\par
	\begin{tikzpicture}
		\node at (0,-20){\large{$E-E_F$}};
	\end{tikzpicture}
	\caption{Calculated total DOS of monolayer
	({\subref{fig:tc_cal_u}}) \ce{Ti3C2},
	({\subref{fig:tzc_cal_u}}) \ce{(Ti_{.97}Zr_{.03})3C2},
	({\subref{fig:tzc2_cal_u}}) \ce{(Ti_{.93}Zr_{.07})3C2},
	({\subref{fig:tcn_cal_u}}) \ce{Ti3CN},
	({\subref{fig:tzcn_cal_u}}) \ce{(Ti_{.97}Zr_{.03})3CN}, and
	({\subref{fig:tzcn2_cal_u}}) \ce{(Ti_{.93}Zr_{.07})3CN}
	using Hubbard parameter $U$($U = 3.0$ for Ti and $U = 2.85$ for Zr).}
	\label{fig:monolayer_dos}
\end{figure*}

We systematically investigated the electronic structure of pristine and \ce{Zr}-doped \ce{Ti3C2} to elucidate the correlation
between their electronic properties and catalytic performance in the hydrogen evolution reaction (HER). Pristine \ce{Ti3C2}, a
well-studied MXene, exhibits high electrical conductivity due to its metallic nature, characterized by a substantial density of
states (DOS) near the Fermi level, as shown in \figref{fig:tc_cal_u}.
The layered structure of \ce{Ti3C2} helps efficient electron mobility, a critical factor for enhancing catalytic reactions.
Likewise, \ce{Ti3CN}, which has mixed carbon and nitrogen
occupancy at the \ce{X}-sites, demonstrates comparable electronic behavior (\figref{fig:tcn_cal_u}). These results are consistent
with earlier DFT studies that have highlighted the metallic character of pristine MXenes and their capacity for electron transfer
in various catalytic contexts \cite{MOHRDARGHAEMMAGHAMI_2018,ENYASHIN_2013}.
Introducing \ce{Zr} atoms into the MXene structure at 3\% and 7\% concentrations induces notable changes in the electronic
structure. The DOS of the Zr-doped MXenes (Figure
\ref{fig:monolayer_dos}(\subref{fig:tzc_cal_u},\subref{fig:tzc2_cal_u},\subref{fig:tzcn_cal_u},\subref{fig:tzcn2_cal_u}))
shows an increase in electronic states near the Fermi level, particularly between -0.5 eV and -3.0 eV. This increase is
attributed to \ce{Zr}’s larger electron affinity compared to \ce{Ti}, resulting in enhanced electron donation. The higher DOS at
the Fermi level for the \ce{Zr}-doped structures indicates a greater availability of electrons, which is critical for promoting
catalytic reactions. These additional states effectively lower the energy barrier for electron transfer to adsorbed hydrogen ions
(\ce{H^+}), thereby facilitating HER \cite{Kibsgaard_2012}.


\begin{table}[h!]
	\centering
	\begin{tabular}{lrccccc}
		\toprule
		\multirow{2}{*}{{MXenes}} & \ce{E_{form}}                           & $|\Delta G_{\ce{H^*}}|$
		                          & \multicolumn{4}{c}{{Bond lengths(\AA)}}
		\\ \cmidrule{4-7}
		                          & (eV/f.u.)                               & (eV)                               & {Ti-C} & {Zr-C}
		                          & {Ti-N}                                  & {Zr-N}
		\\\midrule
		\ce{Ti3C2}                & -11.80                                  & 1.51$^{(0.927)}$\cite   {Ran_2017} &
		2.09                      & --                                      & --                                 &
		--                                                                                                                                \\
		\ce{Ti3CN}                & -9.24                                   & 1.25$^{(0.770)}$\cite{Jiang_2022}  &
		2.10                      & --                                      & 2.09                               & --
		\\
		\ce{TZ3C}                 & -1.30                                   & 0.35                               & 2.10   & 2.23   & -- &
		--                                                                                                                                \\
		\ce{TZ7C}                 & -1.07                                   & 0.34                               &
		2.11                      & 2.23                                    & --                                 & --
		\\
		\ce{TZ3CN}                & -0.78                                   & 0.19                               & 2.08   & --
		                          & 2.07                                    & 2.25
		\\
		\ce{TZ7CN}                & -1.71                                   & 0.16                               & 2.08   & --
		                          & 2.07                                    & 2.23
		\\ \bottomrule
	\end{tabular}
	\caption{Formation energies (${E_{form}}$), Gibbs free energies \dgh~ for hydrogen adsorption on sufaces,
		and the bond lengths of pristine and Zr-doped MXenes. The value in the brackets are the experimental values obtained by other
		groups. The values differ primarily due to the choice of terminators and the choice PBE potentials.}
	\label{tab:table0}
\end{table}

\begin{figure*}[htpb]
	\centering
	\begin{subfigure}[b]{0.24\textwidth}
		\centering
		\includegraphics[width=\textwidth]{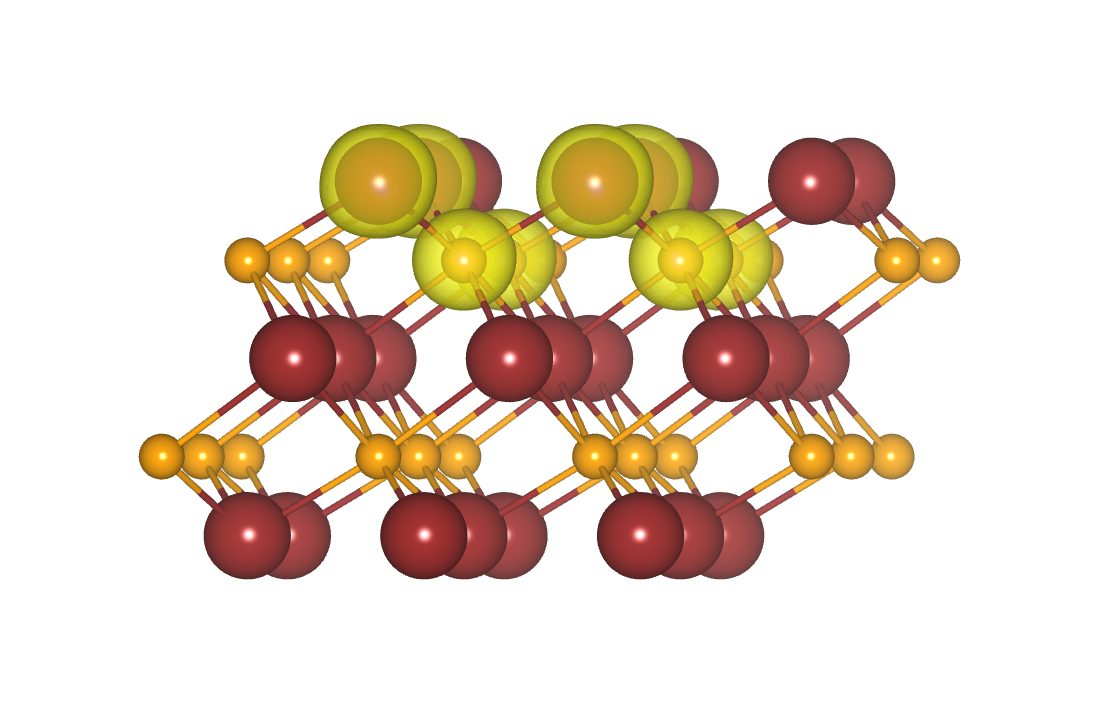}
		\caption{}
		\label{fig:chg_tc}
	\end{subfigure}
	\begin{subfigure}[b]{0.24\textwidth}
		\centering
		\includegraphics[width=\textwidth]{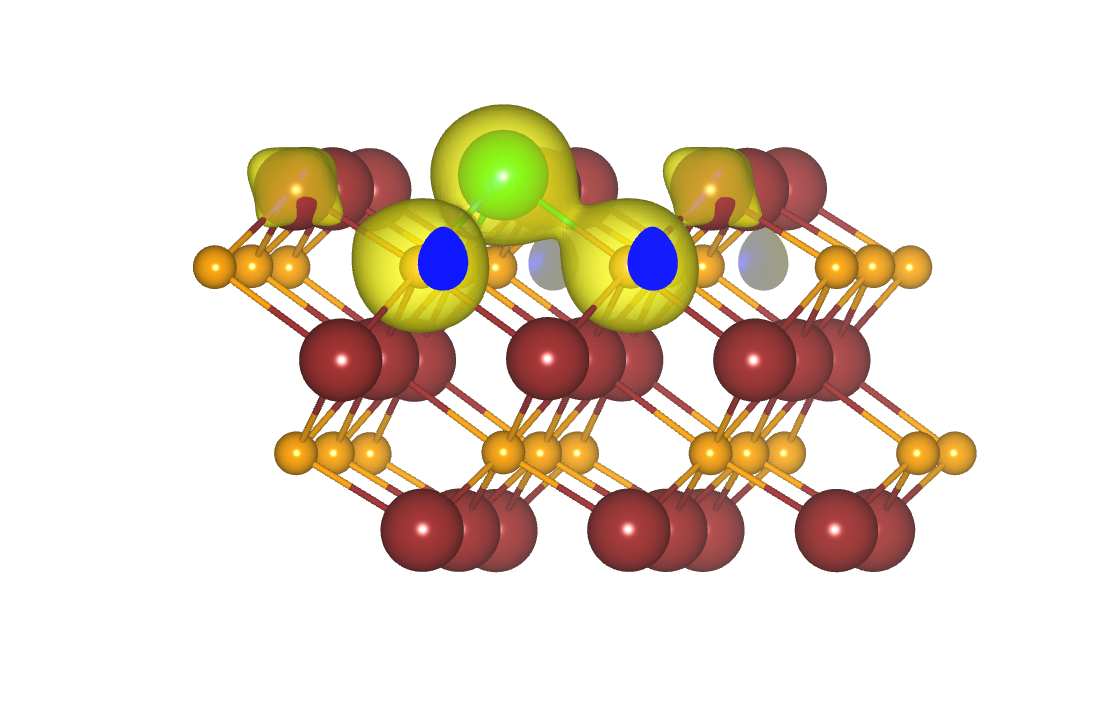}
		\caption{}
		\label{fig:chg_tzc_1}
	\end{subfigure}
	\centering
	\centering
	\begin{subfigure}[b]{0.24\textwidth}
		\centering
		\includegraphics[width=\textwidth]{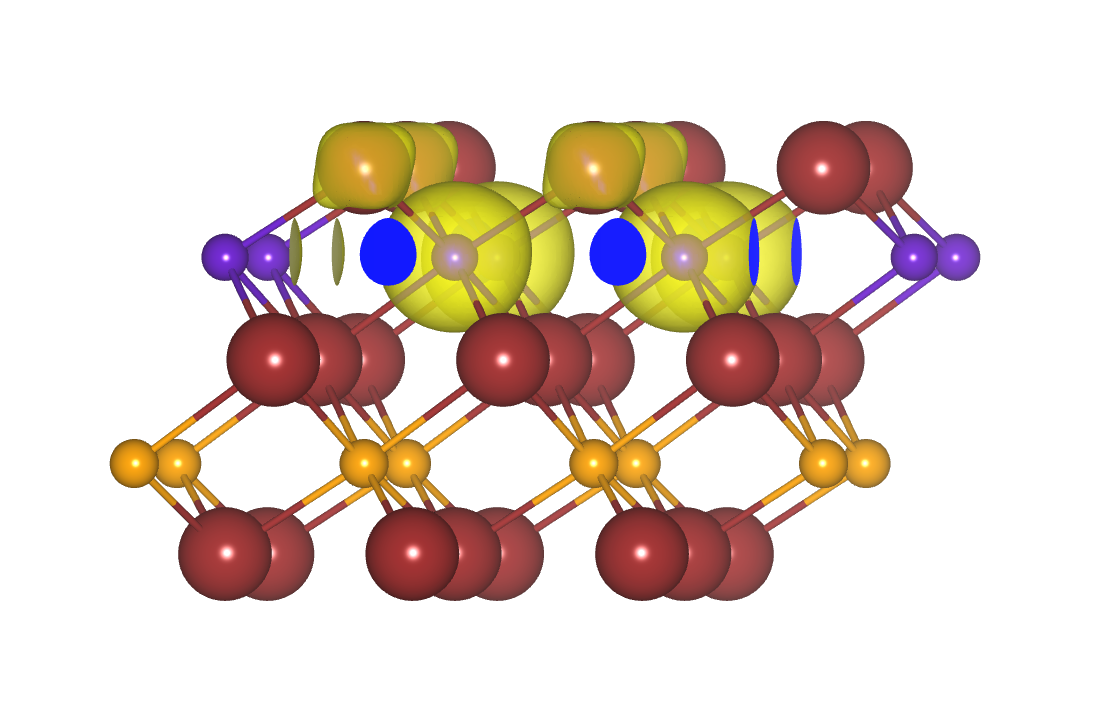}
		\caption{}
		\label{fig:chg_tcn}
	\end{subfigure}
	\begin{subfigure}[b]{0.24\textwidth}
		\centering
		\includegraphics[width=\textwidth]{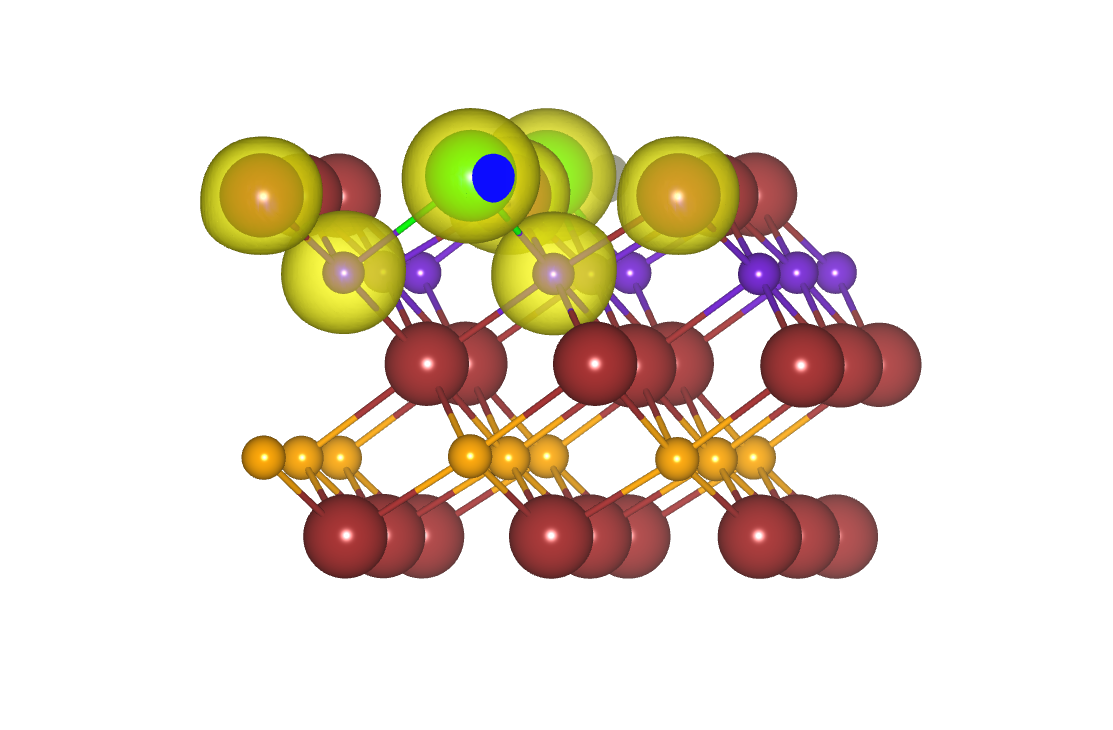}
		\caption{}
		\label{fig:chg_tzcn_2}
	\end{subfigure}
	\caption{Charge density distribution in (\subref{fig:chg_tc}) \ce{Ti3C2};
		(\subref{fig:chg_tzc_1})	\ce{TZ3C};
		(\subref{fig:chg_tcn}) \ce{Ti3CN}
		(\subref{fig:chg_tzcn_2}) \ce{TZ7CN}.
		Ti, Zr, C and N atoms are depicted as maroon, green, orange and violet balls, respectively.}
	\label{fig:chg}
\end{figure*}

\figref{fig:chg} illustrates the Bader charge analysis for four MXene systems. The pristine \ce{Ti3C2}(\figref{fig:chg_tc}) shows
low charge accumulation of \ce{Ti} and \ce{C} atoms.  In contrast, Zr doping, where \dgh~is reduced to $\approx -0.1$ eV,
accumulates large charge on the \ce{Zr} and \ce{N} atoms, as shown in \ce{TZ3C}(\figref{fig:chg_tzc_1}) and \ce{TZ7CN}
(\figref{fig:chg_tzcn_2}).  \ce{Ti3CN},\ce{TZ3C}  and \ce{TZ3CN} shows the similar trend of higher charge accumulation in \ce{Zr}
and \ce{N} compared to \ce{Ti} and \ce{C} and results in lower \dgh.

The enhanced catalytic performance of Zr-doped MXenes can be attributed to the synergistic role of Zr and N. Zr, with its higher electron affinity compared to Ti, facilitates greater charge redistribution within the lattice, increasing surface electron availability crucial for HER. N, being more electronegative than C, stabilizes hydrogen adsorption by modifying the local electronic environment. This combination fine-tunes the Gibbs free energy and provides more favorable adsorption sites for H* intermediates.

In summary, our results reveal that Zr-doping of \ce{Ti3C2} and \ce{Ti3CN} MXenes significantly enhances their catalytic
performance for HER by modifying the electronic structure and reducing the Gibbs free energy of hydrogen adsorption. The improved
electron mobility and increased surface electron density, along with the lowered work function, contribute to the superior
catalytic efficiency of the Zr-doped MXenes. These findings highlight the potential of Zr-doped MXenes as promising candidates
for sustainable hydrogen production, rivalling the performance of traditional platinum-based catalysts.

\section{Conclusion}
Our study demonstrates the significant catalytic potential of Zr-doped Ti-based MXenes for the hydrogen evolution reaction (HER).
Specifically, the Zr-doped single-layered MXenes \ce{(Ti_{.97}Zr_{.03})3C2} and \ce{(Ti_{.97}Zr_{.03})3CN} exhibit \(\Delta
G_{\ce{H^*}}\) values around 0.3 eV, while higher Zr concentrations, as seen in \ce{(Ti_{.93}Zr_{.07})3C2} and
\ce{(Ti_{.93}Zr_{.07})3CN}, result in optimal values of 0.18 and 0.16 eV, respectively. These near-zero \(\Delta G_{\ce{H^*}}\)
values indicate that Zr-doped MXenes, particularly at higher doping levels, have excellent hydrogen adsorption properties,
approaching the catalytic efficiency of platinum-based catalysts.

Our density of states (DOS) analysis confirms that Zr doping enhances the metallic character and increases the electron density
near the Fermi level, which is pivotal for efficient electron transfer during HER. The improved electronic properties directly
translate into enhanced catalytic activity, as demonstrated by the reduction in \(\Delta G_{\ce{H^*}}\) and lower work functions
observed in the doped structures. This enhanced electron mobility and surface electron density, driven by Zr's introduction,
significantly reduces the energy barrier for hydrogen adsorption, thereby facilitating HER.

An important distinction in our study is the focus on unterminated MXenes, a departure from the majority of existing literature,
which primarily investigates MXenes with surface terminations such as \ce{-O}, \ce{-OH}, or \ce{-F}. While surface terminations
are often considered to stabilize MXenes and enhance their chemical reactivity, they introduce complexity that can obscure the
intrinsic properties of the material. By studying unterminated MXenes, we isolate the electronic and catalytic
behaviors of the pristine 2D structures, providing a clearer understanding of their core material properties. This approach
allows for an evaluation of the impact of \ce{Zr} doping on the electronic structure and catalytic activity, particularly
in the context of hydrogen evolution reactions (HER). Furthermore, insights gained from unterminated MXenes form a crucial
baseline for future studies, facilitating a deeper comprehension of how surface functionalization affects performance.  Thus, our
study offers a more fundamental exploration of MXene chemistry, which is essential for designing next-generation HER catalysts.

These findings underscore the potential of Zr-doped Ti-based MXenes as highly efficient and cost-effective alternatives to noble
metal catalysts for hydrogen production. The ability to tailor the electronic structure through controlled doping opens new
pathways for optimizing 2D MXenes for energy conversion technologies. This study paves the way for experimental validation and
further exploration of Zr-doped MXenes in sustainable hydrogen production via water splitting, offering a promising avenue for
green energy applications. Future work could focus on the experimental validation of Zr-doped Ti-based MXenes to confirm the
theoretical predictions. Exploring other dopants, their synergetic effects, and the influence of surface terminations under
different environmental conditions would provide deeper insights into optimizing HER activity. Additionally, integrating machine
learning techniques to identify optimal compositions could accelerate the discovery of efficient catalysts for sustainable
hydrogen production.

\section*{Acknowledgments}
\label{sec:acknowledgement}
We express our sincere gratitude to the high performance computing center (HPCC), SRMIST and the Department of Physics
and Nanotechnology for their support of the computational facility. We are appreciative of the computational resource provided by
the National Param supercomputing facility (NPSF, CDAC), Government of India. We also express our heartiest gratitude to Selective Excellence Research Initiative (SERI), SRMIST for their support in our research work.
\bibliography{draftjpc}

\end{document}